\documentclass[aps,prl,twocolumn,superscriptaddress]{revtex4-2}

\usepackage[compat=1.0.0]{tikz-feynman}
\usepackage{amssymb}
\usepackage{multirow}
\usepackage{bm}
\usepackage{amsmath}
\usepackage{graphicx}
\usepackage{epstopdf}
\usepackage{subfigure}
\usepackage{epsfig}
\usepackage{amsfonts}
\usepackage{mathrsfs}
\usepackage[toc,page,title,titletoc,header]{appendix}
\usepackage[colorlinks,linkcolor=blue,citecolor=blue,anchorcolor=blue]{hyperref}
\usepackage{dsfont,amsthm,amsbsy}
\usepackage{cancel}
\usepackage{ulem}

\usepackage{physics}
\usepackage{enumerate} 
\usepackage[ruled]{algorithm2e} 
\usepackage[framemethod=tikz]{mdframed}

\mdfdefinestyle{question}{
	innertopmargin=1.2\baselineskip,
	innerbottommargin=0.8\baselineskip,
	roundcorner=5pt,
	nobreak,
	singleextra={%
		\draw(P-|O)node[xshift=1em,anchor=west,fill=white,draw,rounded corners=5pt]{%
		Question \theQuestion\questionTitle};
	},
}

\newcounter{Question} 

\mdfdefinestyle{warning}{
	topline=false, bottomline=false,
	leftline=false, rightline=false,
	nobreak,
	singleextra={%
		\draw(P-|O)++(-0.5em,0)node(tmp1){};
		\draw(P-|O)++(0.5em,0)node(tmp2){};
		\fill[black,rotate around={45:(P-|O)}](tmp1)rectangle(tmp2);
		\node at(P-|O){\color{white}\scriptsize\bf !};
		\draw[very thick](P-|O)++(0,-1em)--(O);
	}
}

\mdfdefinestyle{info}{%
	topline=false, bottomline=false,
	leftline=false, rightline=false,
	nobreak,
	singleextra={%
		\fill[black](P-|O)circle[radius=0.4em];
		\node at(P-|O){\color{white}\scriptsize\bf i};
		\draw[very thick](P-|O)++(0,-0.8em)--(O);
	}
}

\usepackage{dsfont}

\allowdisplaybreaks

\newcommand{\be}{\begin{equation}}
\newcommand{\ee}{\end{equation}}
\newenvironment{eqs}
{\begin{equation} \begin{aligned}}
{\end{aligned} \end{equation} }
\newcommand{\bal}{\begin{eqs}}
\newcommand{\eal}{\end{eqs}}

\newcommand{\Ueph}{U_{e\text{-ph}}}

\begin{document}

\title{Semiclassical theory of bipolaronic superconductivity in a bond-modulated electron-phonon model}
\author{Kyung-Su~Kim}
\email{kyungsu@stanford.edu}
\affiliation{Department of Physics, Stanford University, Stanford, California 93405, USA}
\author{Zhaoyu~Han}
\email{zyhan@stanford.edu}
\affiliation{Department of Physics, Stanford University, Stanford, California 93405, USA}
\author{John~Sous}
\email{jsous@ucsd.edu}
\affiliation{Department of Physics, Stanford University, Stanford, California 93405, USA}
\affiliation{Department of Chemistry and Biochemistry, University of California, San Diego,  La Jolla, California 92093, USA}

\date{\today}

\begin{abstract}
We analyze the transition temperature $T_c$ of bipolaronic superconductivity in a bond Su-Schrieffer-Heeger (bond-SSH) model---also known as a bond Peierls model---where the electron hoppings are modulated by bond phonons. Using a semiclassical instanton approximation justifiable in the adiabatic limit of slow phonons, we find that the bipolaron mass is only weakly enhanced, in contrast to the typical large mass enhancement found in standard (Holstein) electron-phonon models. Specifically, in the strong coupling limit, the bipolarons can freely slide within a degenerate manifold rather than become self-trapped. A gas of these bipolarons can undergo a superfluid transition at a critical temperature for which we obtain an upper bound. We find that this bound is exponentially larger than that in the Holstein model. Our study provides an analytical understanding of the mechanism behind the high-$T_c$ bipolaronic superconductivity numerically observed in [Phys. Rev. X 13, 011010 (2023)].  
\end{abstract}

\maketitle

The quest for higher critical transition-temperature ($T_c$) superconductivity has been of extensive interest, because of both fundamental theoretical importance and practical real-world applications~\cite{zhou2021high}.
It is known that the conventional Bardeen-Cooper-Schrieffer (BCS) mechanism of superconductivity leads to a relatively low value of $T_c$ due to an exponential suppression
in the regime of weak electron-phonon coupling strength (characterized by an energy scale $\Ueph$). On the other hand, the value of $T_c$ is also believed to be suppressed in the strong-coupling regime (large $\Ueph$) due to the formation of bipolarons with large mass~\cite{ChakravertyBipolaron, chakraverty1987squeezed, bonca2000mobile,alexandrov2001breakdown,esterlis2018breakdown,esterlis2019pseudogap}. This suppression of $T_c$ in both the weak and strong coupling limits was used to obtain an empirical bound on its value: $k_BT_c \lesssim 0.1 \cdot \hbar \omega$~\cite{ChakravertyBipolaron, CohenBounds, esterlis2018bound, hofmann2022boundTc,Chubukov} (where $\omega$ is the characteristic phonon frequency), the maximum of which is believed to arise at the ``sweet spot'' of intermediate $\Ueph$.

The supposed ``suppression'' of $T_c$ in the strong-coupling regime is based on the reasoning that the bipolarons always suffer from severe mass enhancement, which renders their dynamics rather slow. 
Such an expectation was challenged in recent numerical studies of bond Peierls/Su-Schrieffer-Heeger (bond-SSH) model~\cite{zhang2023bipolaron,SousCoulomb}, which reported unusually weak enhancement of the bipolaron mass~\cite{Sous2018} and a concomitant large value of $T_c$ that exceeds the aforementioned empirical bound. 
In this Letter, we aim to give an explanation for this unusual enhancement of $T_c$ with a controllable analysis in the adiabatic limit of slow phonons, which is of relevance to real materials.  Our main findings are as follows.
(i) We show that the bipolaron mass remains unusually small. (ii) We unveil a semiclassical picture behind the very weak mass enhancement: precisely, we find that in the limit of zero bare hopping $t$, there exists a degenerate manifold of bipolaron states within which a bipolaron can slide freely without any energy cost. The inclusion of a small bare hopping lifts the degeneracy and leads to a finite but small energy barrier, resulting in a weak mass enhancement. Thus, surprisingly, the suppression of $T_c$ is least severe in the ``atomic limit'' ($t=0$) corresponding to a trivial flat band. (iii) We estimate the bipolaron mass and radius to obtain an empirical bound on the value of $T_c$ of the bipolaronic superconductivity, showing that it can be much larger than that in the standard (Holstein) models. Our results provide an analytic understanding of recent numerical results on the bipolaronic high-$T_c$ superconductivity found in the bond-SSH model~\cite{zhang2023bipolaron,SousCoulomb,wang2022robust,tanjaroon2023comparative}.
In a broader view, we hope our work provides another concrete example of various unexpected phenomena driven by strong bond-modulated electron-phonon coupling~\cite{Sous2020Fracton2,CarboneBond,BoSSH2D,PhysRevLett.127.247203,ScaletterSSHU,PhysRevB.106.L081115,PhysRevB.105.085151,PhysRevLett.130.186404,gotz2023phases,costa2023comparative}.

{\em Model and methods.} We study the bond-SSH model on a two-dimensional square lattice described by the Hamiltonian:
\begin{align}
    \hat{H} &=\hat{H}_e + \hat{H}_\text{ph} \\
    \hat{H}_e & = -\sum_{\langle ij\rangle,\sigma} \left(t+\alpha \hat{X}_{\langle ij \rangle}\right)  \left(\hat{c}_{i\sigma}^\dagger \hat{c}_{j\sigma} + {\rm H.c.} \right) \\
    \hat{H}_\text{ph} &=\sum_{\langle ij\rangle} \left(\frac{K\hat{X}^2_{\langle ij \rangle}}{2}  + \frac{\hat{P}^2_{\langle ij \rangle}}{2M} \right)
\end{align}
where the coordinates of the phonons $\hat X_{\left \langle ij \right >}$  live on nearest-neighbor bonds $\langle ij \rangle$. Each local phonon mode has a frequency $\omega_0 \equiv \sqrt{K/M}$. Besides $t$  and $\hbar\omega_0$, there is only one {\it independent} energy scale in this problem, $\Ueph \equiv \alpha^2/K$, which characterizes the strength of the electron-phonon coupling~\footnote{This can be most easily seen by rescaling the phonon coordinates $\hat{X}\rightarrow \frac{\alpha}{K} \hat{X}$ in the Hamiltonian, rendering them dimensionless. }. We set the lattice constant $a=1$. Without loss of generality, we will consider the $t>0$ case throughout the paper.

\begin{figure}[!t]
    \centering
    {\includegraphics[width=\linewidth]{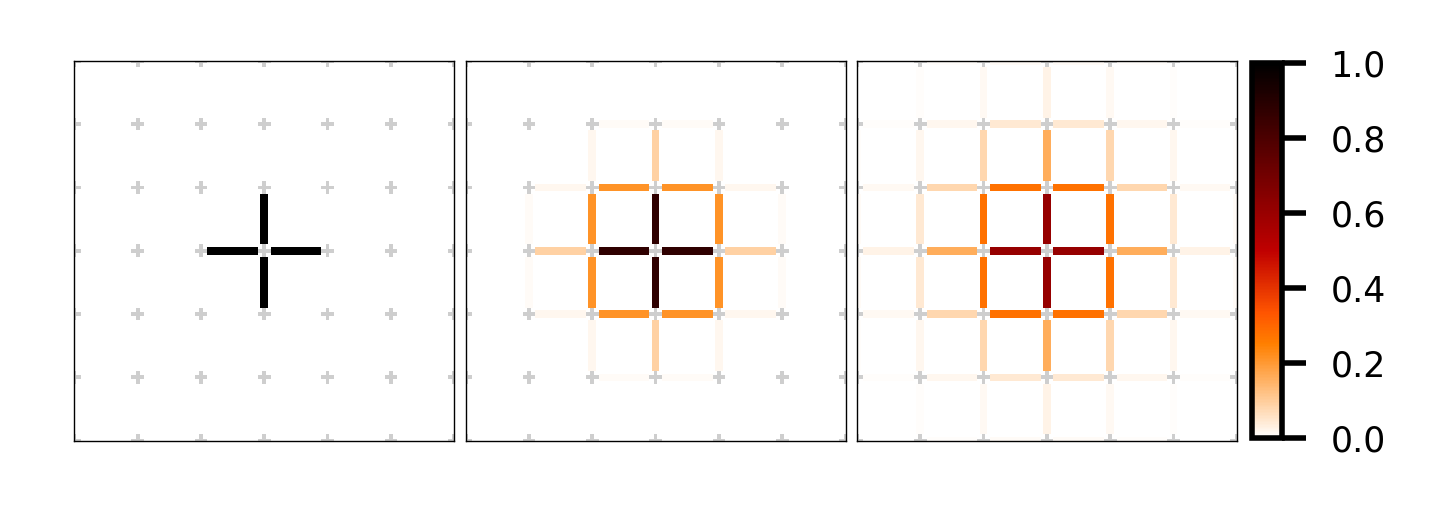}}
    
    \caption{The phonon configuration $\{X_{\langle ij\rangle}\}$ (measured in units of $\alpha/K$) around a static bipolaron ($\omega_0=0$) for different values of $t/\Ueph=$ 0.01 (left), 0.35 (middle) and 0.7 (right).}\label{fig: phonon config}
\end{figure}

We analyze the problem in the adiabatic limit $\hbar\omega_0/\Ueph \rightarrow 0$ for dilute electron concentrations, where two electrons form a bound bipolaron and move around with the help of phonons. As we will show, this situation is effectively described by a gas of repulsively interacting bosons, which undergo a superfluid transition at a temperature $T_c$~\cite{fisherHohenberg1988, prokof2001critical, pilati2008critical}. We will first solve the problem in the static (classical) limit where $\omega_0 =0$ ($M = \infty$). In such a case, one can obtain the electronic ground state energy in the two-electron sector, 
\begin{align}
\label{eq:V_eff}
    \hat{V}_\text{eff} \left ( \{\hat{X}_{\langle ij\rangle}\}\right)  &= 2 \epsilon_0 \left ( \{\hat{X}_{\langle ij\rangle}\}\right) +  \sum_{\langle ij\rangle} \frac{K\hat{X}^2_{\langle ij \rangle}}{2},
\end{align}
where 2$\epsilon_0 \left ( \{\hat{X}_{\langle ij\rangle}\}\right)$ is the ground state energy of $\hat H_{e}$ in the two-electron sector for a given phonon configuration $\{\hat{X}_{\langle ij\rangle}\}$. We then minimize $\hat V_{\rm eff}$ with respect to $\{X_{\left \langle ij \right >}\}$ to obtain the bipolaron bound state with the smallest energy. Such bipolaron states, localized around different sites, are classically degenerate. 
For a small but finite phonon frequency $\omega_0 > 0$, these bipolarons acquire quantum dynamics  described by the  effective Hamiltonian
\begin{align}
   \hat{H}_\text{eff} &=\hat{V}_\text{eff} +  \sum_{\langle ij\rangle}  \frac{\hat{P}^2_{\langle ij \rangle}}{2M}. 
\end{align}
We will use a semiclassical instanton approximation to calculate the effective hopping of the bipolaron, which in turn gives an expression for the bipolaron effective mass $m^*$.
From this, we estimate the value of the critical temperature $T_c$ of superconductivity of a gas of these bipolarons, which is  given by~\cite{fisherHohenberg1988, prokof2001critical, pilati2008critical}:
\begin{align}
\label{eq:T_c}
    T_c \approx C \frac{\hbar^2 \rho}{m^*k_B},
\end{align}
where $C \equiv \frac{2 \pi}{\ln(\frac {380} {4\pi})} \approx 1.84$. The maximum possible $T_c$ is achieved for the density of electrons at which bipolarons start to overlap, $\rho = \frac 1 {\pi R^2}$, where $R$ is the linear size of the bipolaron. We note that this $T_c$ is not determined by the pairing energy scale as in the case of BCS or Migdal-Eliashberg theory~\cite{Chubukov}, but rather by the phase coherence energy scale within the Bose-Einstein Condensation (BEC) picture of preformed pairs.

{\em Static limit.}
In the static (classical) limit where the phonon kinetic energy is neglected ($\omega_0=0$), we perform numerical optimization ~\cite{optim} to find the phonon configurations that minimize the effective potential $V_\text{eff}(\{X_{\langle ij\rangle}\})$. 
We first consider the case corresponding to two electrons in the system. 
For the special case when $t=0$, we obtain a manifold of degenerate bipolaron configurations localized around each site $i$. Such a manifold is parametrized by four phonon coordinates $X_{\left \langle ii_k \right \rangle}$ ($k=1,2,3,4$), where $i_1, ..., i_4$ are four nearest-neighbor sites connected to the site $i$ (other phonon coordinates are zero).
Minimizing the electronic energy within such phonon configurations, one obtains 
\begin{align}
    V_\text{eff}\left(\{X_{\langle ij\rangle}\}\right)= -2 \alpha \left [ \sum _{k=1}^4 X_{\left \langle ii_k \right \rangle}^2 \right ]^{\frac 1 2 } + \frac{K}{2}\sum _{k=1}^4 X_{\left \langle ii_k \right \rangle}^2,
\end{align}
with the electronic eigenfunction $ \left | \Psi^0_{i,\sigma  = \uparrow,\downarrow} (\{X \}) \right \rangle $ with support on five sites $i,$ $i_1,$ $i_2,$ $i_3$ and  $i_4$. $V_\text{eff}\left(\{X_{\langle ij\rangle}\}\right)$ is minimized when 
\begin{align}
\label{eq:bp manifold}
    \left [ \sum _{k=1}^4 X_{\left\langle i i_k \right \rangle}^2 \right ]^{\frac 1 2 } = \frac{2\alpha}{K}
    \end{align} 
    with energy $-2 U_{e\text{-ph}}$.
This degenerate manifold of bipolaron configurations means that the bipolaron can ``slide'' freely from one site to the other by deforming its configuration within the manifold. 
This is responsible for the small bipolaron mass in the strong-coupling limit $t/U_{e\text{-ph}} \lesssim 1$ as we will see later.

\begin{figure}[t!]
    \centering
    {\includegraphics[width=\linewidth]{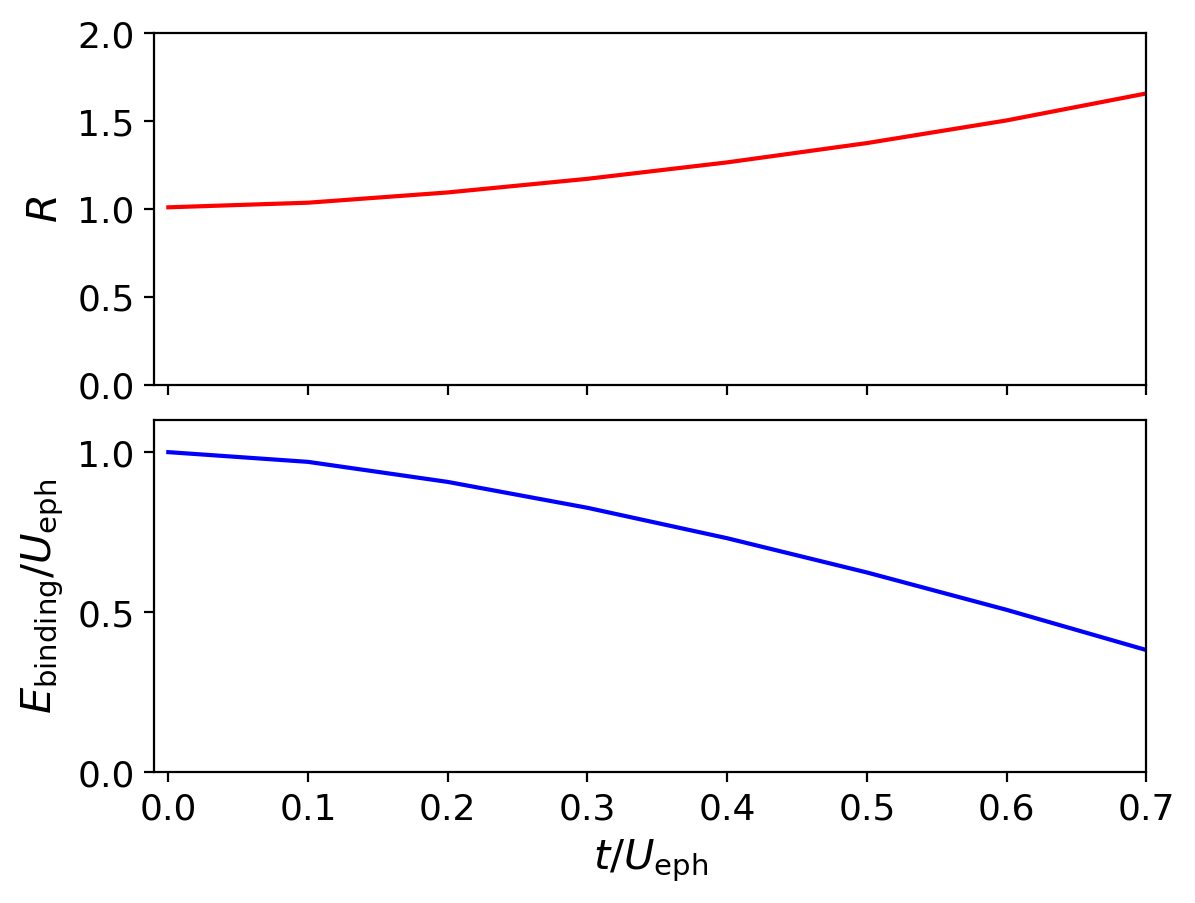}}
    
    \caption{The properties of bipolarons in the static limit $\omega_0=0$.  We plot the effective radius $R$ of the bipolaron (Eq.~\ref{eq: radius}) in the upper panel and the binding energy $E_\text{binding}$ (Eq. \ref{eq:binding energy}) in units of $\Ueph$ in the lower panel as functions of the dimensionless constant $t/\Ueph$. 
    }\label{fig: static properties}
\end{figure}

\begin{figure*}[t!]
    \centering
    {\includegraphics[width=\linewidth]{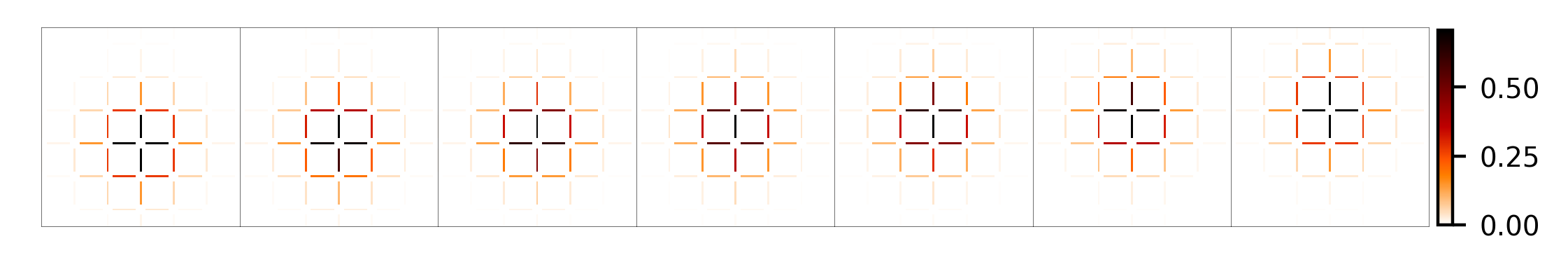}}
    
    \caption{Snapshots of an imaginary time tunneling trajectory at $7$ representative time slices connecting the initial (left-most panel) and the final (right-most panel) bipolaron configurations. 
    The color encodes the configuration of the phonon displacement $\{X_{\langle ij\rangle}\}$  (measured in unit of $\alpha/K$). This simulation is performed for $t/\Ueph = 0.6$. }\label{fig: trajectory}
\end{figure*}

Including a $t > 0$ term determines the unique lowest energy bipolaron configuration.  In the small-$t$ limit, first-order perturbation theory gives an energy correction for each configuration $\{X_{\left \langle ii_k \right \rangle} \}$ within the manifold, splitting the energies of different bipolaron configurations (\ref{eq:bp manifold}):
\begin{align}
    \Delta E &= -2t \left \langle \Psi^e_{i,\uparrow}  \right | \sum_{\left \langle ij \right \rangle} c_{i,\uparrow}^{\dagger}c_{j,\uparrow} \left | \Psi^e_{i,\uparrow}  \right \rangle  + O(t^2)
    \nonumber \\
    &=  -  \frac t {\alpha/K} \sum_{k=1}^4 X_{\left \langle ii_k \right \rangle} \geq - 4t.
\end{align}
Here, the energy minimum (equality) is attained for the uniform ``cross'' configuration $X_{\left \langle ii_k \right \rangle} = \alpha/K$ (see the left panel of Fig.~\ref{fig: phonon config}).
Note that bipolarons centered around different sites are related by a lattice translation and are thus classically degenerate. 

With increasing $t/U_{e\text{-ph}}$, bipolarons become more and more extended (see Fig.~\ref{fig: phonon config}). To better quantify this observation, we compute the inverse participation ratio (IPR$\equiv \sum_i |\Psi_{i,\sigma}^0|^4$) of the electronic ground state $|\Psi^0_{i,\sigma}\rangle$ of $\hat{H}_e$ within the optimal $\{X_{\langle ij \rangle}\}$ and use this quantity to define an effective radius of the bipolaron:
\begin{align}\label{eq: radius}
    R \equiv \frac{1}{\sqrt{\pi (\mathrm{IPR}) }}.
\end{align} 
Another meaningful quantity is the energy gain associated with the formation of a bipolaron, defined as
\begin{align}
\label{eq:binding energy}
    E_{\rm binding} = 2 E_{\rm polaron} - E_{\rm bipolaron},
\end{align}
where $E_{\rm polaron}$ is the ground state energy in the single-electron sector. We plot these two quantities as a function of $t/\Ueph$ in Fig.~\ref{fig: static properties}~\footnote{Due to limitations of numerical resources, we only calculate the properties of bipolarons  for the range $0<t/\Ueph\leq0.7$ 
As $t/\Ueph$ grows, our numerics do not converge well for systems larger than $70\times 70$.  We also note that it is likely that static bipolarons become fully delocalized throughout the entire system at some critical value of $t/\Ueph$ beyond which our method breaks down.}. Interestingly, we find that the size of the bipolaron increases only slowly as $t/\Ueph$ is increased.

Importantly, when considering a finite, but dilute density of electrons in the system, we find that they always tend to repel each other and phase separation does not preempt the superconducting phase~\cite{NoceraPS}.

\begin{figure}[t!]
    \centering
    {\includegraphics[width=\linewidth]{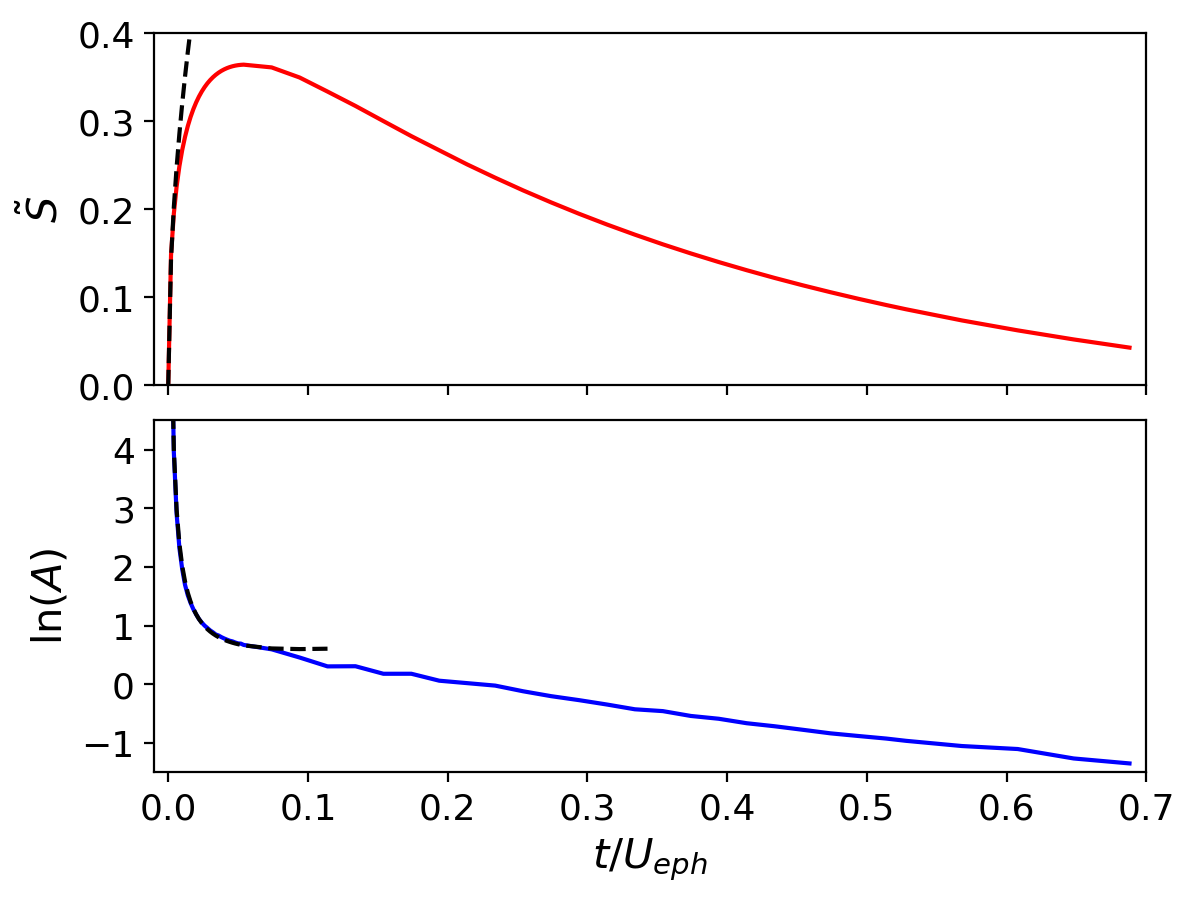}}
    
    \caption{The energetics of bipolaron hopping. We plot the normalized classical action of the nearest-neighbor tunneling process, $\tilde{S} \equiv \frac{\hbar \omega_0}{\Ueph} S_{\rm inst}$, in the upper panel,  and the leading-order fluctuational correction to it, $\ln(A)$, in the lower panel. In the small $t/\Ueph$ limit, we find that the numerical results are well fitted by $\tilde{S} \approx 3.2 \sqrt{t/\Ueph}$ (upper panel) and $ \ln(A) \approx \ln(2.4) + \frac{3}{4} \ln\frac{t}{\Ueph} + 0.46 \sqrt{\frac{\Ueph}{t}}$ (lower panel) which are drawn as dashed lines.} 
    \label{fig: energetics}
\end{figure}

{\em Adiabatic limit.}
When a finite but small phonon kinetic energy is considered (i.e., $\hbar \omega_0/\Ueph\rightarrow 0$), the classically degenerate bipolaron configurations centered at different sites become connected by quantum tunneling. The bipolaron hopping matrix element can be computed using the standard semiclassical  instanton approximation  in the two-electron sector,
\begin{align}
\label{eq:instanton}
    &\tau_{\rm eff} = \hbar \omega_0\ A \sqrt{\frac{ S_{\rm inst}}{2\pi}} \mathrm{e}^{-  S_{\rm inst}  },
\end{align}
where $S_{\rm inst}$ is the action of the semiclassical path that connects the initial ($\mathrm{init}$) and the final ($\mathrm{final}$) bipolaron configurations, 
\begin{align}
     S_{\rm inst} \equiv
    \frac{\Ueph}{\hbar \omega_0} \tilde S={\rm min} \int_{X_{\rm init}}^{X_{\rm final}} dX \sqrt{2M \left [ V_{\rm eff}(X) - E_0 \right ]},
\end{align}
where $E_0 \equiv V_{\rm eff}(X_{\rm init})= V_{\rm eff}(X_{\rm final})$ is the classical energy of a bipolaron (see Fig.~\ref{fig: trajectory} for an exemplary trajectory). $A$ is the ``fluctuation determinant," the sub-leading fluctuational correction, that captures Gaussian fluctuation around the semiclassical tunneling path 
\begin{align}
    A = \left [ \frac{ \omega_0^2 \  \widetilde{\rm det} (-\partial_{\tau}^2 + \frac 1 M V''_{\rm eff}[ X_{\rm cl}(\tau)])}{{\rm det} (-\partial_{\tau}^2 + \frac 1 M V''_{\rm eff}[X_{\rm init}])} \right]^{-\frac 1 2}.
\end{align}
Here, $\widetilde{\rm det}$ is the determinant without the zero eigenvalue~\cite{altlandSimons2010, coleman1988aspects, zinnJustinQFT}, $ X_{\rm cl}(\tau)$ is the instanton trajectory and the double prime denotes the second functional-derivative of $V_{\rm eff}[X(\tau)]$. We only consider a nearest-neighbor hopping matrix element, since further-ranged tunneling processes will be exponentially smaller in the $\omega_0 \rightarrow 0$ limit.
We numerically evaluate the normalized instanton action $\tilde S$ and  $A$ as a function of $t/\Ueph$ in Fig.~\ref{fig: energetics}
(See Ref.~\cite{voelker2001disorder, kim2023defects} for details of their calculation). The effective bipolaron mass can in turn be estimated as $m^* = \frac{\hbar^2}{4\tau_{\rm eff}}$. Note that including further-neighbor hoppings of bipolarons will make $m^*$ even smaller.

{\em Empirical bound on $T_c$}. 
The maximum possible value of $T_c$ from Eq. \ref{eq:T_c} arises for the density at  which bipolarons start to overlap $\rho_{\rm max} \sim \frac{1}{\pi R^2}$:
\begin{align}
    \label{eq:max T_c}
     k_B T_c^{\rm max} \sim 
      \frac{\sqrt{\hbar\omega_0 \Ueph} }{R^2}\  A \sqrt{\frac{\tilde S}{2\pi}} \  e^{-\frac{\Ueph}{\hbar\omega_0} \tilde S} .
\end{align}
We find that $\Tilde{S} < 0.35$ which is much smaller than the corresponding quantity in the Holstein model with density coupling $\alpha \sum_i X_i n_i $, $\Tilde{S}^\text{Holstein}=2$~\cite{PhysRevLett.125.167001}. This implies that the value of $T_c$ in the bond model can be exponentially larger than that in the Holstein model in the adiabatic limit $\hbar\omega_0/\Ueph\rightarrow 0$.

In the limit $t/\Ueph \rightarrow 0$, we further obtain the asymptotic behaviors of the action and the fluctuation determinant: $\tilde S \approx  3.2  \sqrt{\frac{t}{\Ueph}} \to 0$ and $ A \approx 2.4 \cdot (\frac{t}{\Ueph})^{3/4} e^{{0.46}\cdot\sqrt{{\frac{\Ueph}{t}}}} \to \infty$, as shown in Fig.~\ref{fig: energetics}. That is, as $t/\Ueph \to 0$, the tunnel barrier between the two neighboring bipolaron configurations tends to zero and the fluctuation determinant diverges. 
This is due to the continuous family of degenerate bipolaron configurations (Eq.~\ref{eq:bp manifold}).
For $t=0$, the two nearest-neighbor ``cross'' bipolaron configurations are connected by a zero tunnel barrier through Eq.~\ref{eq:bp manifold}. For small $t,$ the first order correction in the energy of non-cross configuration is $\Delta E = O(t)$, and hence the tunnel barrier height $\Delta V_{\rm eff} \sim t$.
On the other hand, the tunnel distance in coordinate space $\Delta X \sim \alpha /K$. Therefore,
\begin{align}
    \tilde S \sim \frac{\hbar\omega_0}{\Ueph}\Delta X \sqrt{2M \Delta V_{\rm eff}} \sim \sqrt {\frac t {\Ueph}}.
\end{align}
Moreover,   the instanton duration $\Delta \tau_{\rm inst}$ in  imaginary time becomes longer as $t/\Ueph \to 0$, due to the small tunnel barrier
\begin{align}
    \Delta \tau_{\rm inst} = \int dX \sqrt{\frac{M}{2\Delta  V_{\rm eff}(X) }} 
    \sim \frac{1}{\hbar \omega_0}\cdot \sqrt{\frac{\Ueph}{t}}.
\end{align}
The fluctuation determinant can be estimated as the ratio between the two harmonic oscillator propagators during the interval $\Delta \tau_{\rm inst}$ with the dynamical matrix evaluated for an intermediate (initial) $X_{\rm middle}$ ($X_{\rm init}$) configuration 
\begin{align}
\label{eq:asymptotic of A}
    &A \sim  \left [ \frac{{\rm det} (-\partial_{\tau}^2 + \frac 1 M V''_{\rm eff}[ X_{\rm middle}]) \vert_{(0,\Delta \tau_{\rm inst})}}{{\rm det} (-\partial_{\tau}^2 + \frac 1 M V''_{\rm eff}[X_{\rm init}])\vert_{(0,\Delta \tau_{\rm inst})}} \right]^{-\frac 1 2}
    \nonumber \\ 
    &= \left (\prod_{n} {\frac{\omega^{(n)}_{\rm middle}}{\omega^{(n)}_{\rm init}}}  
    \frac{\sinh( \hbar \omega^{(n)}_{\rm init} \Delta \tau_{\rm inst})}{\sinh( \hbar \omega^{(n)}_{\rm middle} \Delta \tau_{\rm inst})}\right )^{1/2} 
    \nonumber \\
    &\sim  \left (\frac{t}{\Ueph} \right )^{3/4} \exp\left [\# \cdot  \sqrt{\frac{\Ueph}{t}} \right ].
\end{align}
In the first line, $\det(...)\vert_{(0,\Delta \tau_{\rm inst})}$ denotes that the eigenvalues of the differential operator must be obtained with the boundary condition $\delta X(0) = \delta X(\Delta\tau_{\rm inst}) =0,$ where $\delta X(\tau)$ denotes the fluctuation coordinate $X(\tau) = X_{\rm cl}(\tau) + \delta X(\tau)$.
The second line is the exact form of the harmonic oscillator propagator, where $\omega^{(n)}_{\rm init}$ and $\omega^{(n)}_{\rm middle}$ are the square root eigenvalues of the respective dynamical matrices $\frac 1 M V''_{\rm eff}[X_{\rm init}]$ and $\frac 1 M V''_{\rm eff}[X_{\rm middle}]$. In the third line, we used the fact that there are three (six)  nearly zero oscillator frequencies $\omega^{(n)}_{\rm init}$ ($\omega^{(n)}_{\rm middle}$) of order $O\left (\sqrt{\frac{t}{\Ueph}}\omega_0 \right)$, and in general ${\rm Re} \sum_n \left (\omega^{(n)}_{\rm init} -  \omega^{(n)}_{\rm middle} \right ) = O(\omega_0) $. As shown in Fig. \ref{fig: energetics}, the asymptotic behavior extracted from numerical calculation fits well with the asymptotic form Eq. \ref{eq:asymptotic of A}.

These asymptotic behaviors lead to an unconventional scaling of $T_c$ of the bipolaronic superconductor
\begin{align}\label{Tc_Bond}
    T_c \sim  t \sqrt{\frac{\hbar\omega_0}{\Ueph}} \exp[-3.2 \frac{\sqrt{t\Ueph}}{\hbar\omega_0} + 0.46 \sqrt{\frac{\Ueph}{t}} ] 
\end{align}
which is valid when $\hbar\omega_0\ll t \ll \Ueph$ (see below for justification) and is parametrically larger than that obtained in the Holstein model: $  T^\text{Holstein}_c \sim \frac{t^2}{\sqrt{{\Ueph \hbar\omega_0}}} \exp[-\frac{2\Ueph}{\hbar \omega_0}]$~\cite{PhysRevLett.125.167001}.

{\em Validity of the approximations}.
Our theory is based on two approximations, which can be justified as long as certain parameter conditions are satisfied. 

The first  is the adiabatic approximation, i.e., the assumption that electronic states adiabatically follow the instantaneous ground state corresponding to the phonon configurations, which is justified as long as~\cite{PhysRevLett.102.220401}:
\begin{align}
    \frac{\left|\langle \Psi^0(\tau)|
    \partial_\tau \hat{H}_e|\Psi^n(\tau)\rangle \right|}{\Delta_{0n}^2} \ll 1
\end{align}
where $\hat{H}_e(\tau)$ denotes the instantaneous electronic Hamiltonian, $\Psi^n(\tau)$ is its $n$-th eigenstate, and $\Delta_{0n}(\tau)\equiv E_n(\tau) - E_0(\tau)$ is an instantaneous gap between $n$-th excited state and the ground state ($\tau$ here is the imaginary time). To ensure this is satisfied, we note that $\frac{\left|\langle \Psi^0|
\partial_\tau \hat{H}_e|\Psi^n\rangle \right|}{\Delta_{0n}^2(t)} \lesssim \frac{\alpha|\dot{X}|} {\Delta^2_\text{min}}$ where $\Delta_\text{min}$ is the minimum energy gap along the trajectory and $|\dot{X}|$ is the speed associated with the coordinates $\{X_{\langle ij \rangle}\}$. 
The phonon trajectory reaches maximum speed when the coordinates are at the position of the largest energy barrier, and the maximum value of the speed becomes $|\dot{X}|_\text{max} = \sqrt{\frac{2\Delta V_{\rm eff}}{M}}$
where $\Delta V_{\rm eff}$ is the height of the energy barrier. Thus, a sufficient condition for the adiabatic theorem to be valid in our treatment is
\begin{align}
    \frac{\hbar \omega_0}{\Ueph} \ll \frac{\Delta_\text{min}^2}{\sqrt{\Delta V_{\rm eff} \Ueph^3 }}. \label{eq: adiabaticity} 
\end{align}

The second instanton approximation is, technically speaking, a semiclassical approximation of the tunneling event, which is valid when the saddle point contribution dominates the path integral:
\begin{align}
\label{eq:saddle point}
    S_{\rm inst} \gg |\ln(A)|.
\end{align}
For $t/\Ueph \ll 1$, this amounts to the condition $t \gg \hbar\omega_0$, whereas for $t/\Ueph = O(1)$, it requires $\Ueph \gg \hbar\omega_0$ (given that both $\tilde{S}= S_{\rm inst}\hbar\omega_0/\Ueph$ and $\ln(A)$ are $O(1)$). 

In our numerical simulation of the instanton processes, we find that the adiabaticity condition, Eq.~\ref{eq: adiabaticity}, is always satisfied whenever Eq.~\ref{eq:saddle point} is satisfied. 
Therefore, in the semiclassical limit properly defined in Eq.~\ref{eq:saddle point}, our results are {\it asymptotically exact}.

{\em Acknowledgements. } KSK and ZH contributed equally to this work.
We thank Steven A. Kivelson, Hong Yao, and Chaitanya Murthy for the helpful discussions. We are also grateful for insightful discussions during the Polaron Meeting at the Center for Computational Quantum Physics (CCQ) of the Flatiron Institute. KSK acknowledges the hospitality of the Massachusetts Institute of Technology, where this work was completed. KSK acknowledges support from the Department of Energy, Office of Basic Energy Sciences, Division of Materials Sciences and Engineering, under contract DE-AC02-76SF00515. JS acknowledges support from the Gordon and Betty Moore Foundation’s EPiQS Initiative through Grant GBMF8686 at Stanford University. 

\bibliography{ref}

\onecolumngrid
\end{document}